\documentclass[reprint,amsmath,amssymb,aps,prb]{revtex4-2}

\usepackage[colorlinks,citecolor=blue,linkcolor=blue]{hyperref}
\usepackage{graphicx}
\usepackage{epstopdf}
\usepackage{amsmath,bm}
\usepackage{threeparttable}
\usepackage{multirow}
\usepackage{booktabs}
\usepackage{color}

\begin{document}

\title{Topological Phases, Local Magnetic Moments, and Spin Polarization Triggered by $C_{558}$-Line Defects in Graphene}%
\author{Ning-Jing Yang$^{1,2}$}%
\author{Wen-Ti Guo$^{1,2}$}%
\author{Hai Yang$^3$} 
\author{Zhigao Huang$^{1,2}$}%
\author{Jian-Min Zhang$^{1,2}$}%

\email[Corresponding author]{jmzhang@fjnu.edu.cn} 

\affiliation{1 Fujian Provincial Key Laboratory of Quantum Manipulation and New Energy Materials, College of Physics and Energy, Fujian Normal University, Fuzhou 350117, China
}
\affiliation{2 Fujian Provincial Collaborative Innovation Center for Advanced High-Field Superconducting Materials and Engineering, Fuzhou, 350117, China
}
\affiliation{3 School of Physics Science and Technology, Kunming University, Kunming 650214, China}
\date{\today}

\begin{abstract}
We study the electronic properties of a novel topological defect structure for graphene interspersed with $C_{558}$-line defects along the armchair boundary. This system has the topological property of being topologically three-periodic and the type-\uppercase\expandafter{\romannumeral2} Dirac-fermionic character of the embedded topological phase. At the same time, we show computationally that the topological properties of the system are overly dependent on the coupling of this line defect. Using strain engineering to regulate the magnitude of hopping at the defect, the position of the energy level can be easily changed to achieve a topological phase transition. We also discuss the local magnetic moment and the ferromagnetic ground state in the context of line defects, which is the conclusion after considering additional Coulomb interactions. This leads to spin polarization of the whole system. Finally, by modulating the local magnetic moment at the position of the line defect, we achieve a tunable spin quantum conductance in a one-dimensional nanoribbon. Near the Fermi energy level, it also has the property of complete spin polarization. Consequently, spin filtering can be achieved by varying the incident energy of the electrons.
\end{abstract}

\maketitle

\section{Introduction}
Extensive research conducted has led to the recognition of the traditional bandgap topological phase \cite{1,2,3,4,5,6,7,8}, with advancements observed in topological properties across periodically driven Floquet systems, non-Hermitian systems, and traditional wave systems \cite{9,10,11,12,13}. 
Recently, a new topological phase related to defect induction has gained attention \cite{14}, where the term "defects" here refers to topological defect structures such as vortices, dislocations, and grain boundaries \cite{15,16,17}. The embedded topology structure \cite{18,19} is of particular interest, as it involves introducing a defect into a regular system, causing it to undergo a transition from a trivial to a topological state. However, the embedded topology insulator has not been extensively explored in real materials. Armchair graphene nanoribbon exhibits this potential topological property at the boundary \cite{20,21,22}. This suggests that further investigation into the embedded topology structure of real materials may reveal novel topological properties.

On the other hand, some experimental results show that two-dimensional graphene grain boundary structures can spontaneously break time-reversal symmetry, which is caused by local magnetic moments arising at the interface position \cite{23}. This is different from magnetism arising from the loss of atoms \cite{24}. 
To elucidate the magnetism induced by the interface, a recently developed DFT+U+V computational method is employed, which incorporates Coulombic interaction  \cite{25,26,27}. This method takes into account the Hubbard interaction between sites and allows a more accurate optimization of the band structures for irregular structures \cite{28}. Specifically, for two-dimensional carbon-based materials, incorporating Hubbard V enables the attainment of a lower energy ferromagnetic ground state \cite{29}. This method has been successfully applied to the calculation of silicon, black phosphorus, graphene and carbon allotropes \cite{26,28,29}. 
Meanwhile, for graphene, inducing spin-polarized electron transport is challenging and often requires regulatory means \cite{30,31}. This spin-polarized transport property can be used as a quantization spin filter \cite{32,33,34,35}.

\begin{figure*}
	\centering
	\includegraphics[width=13cm]{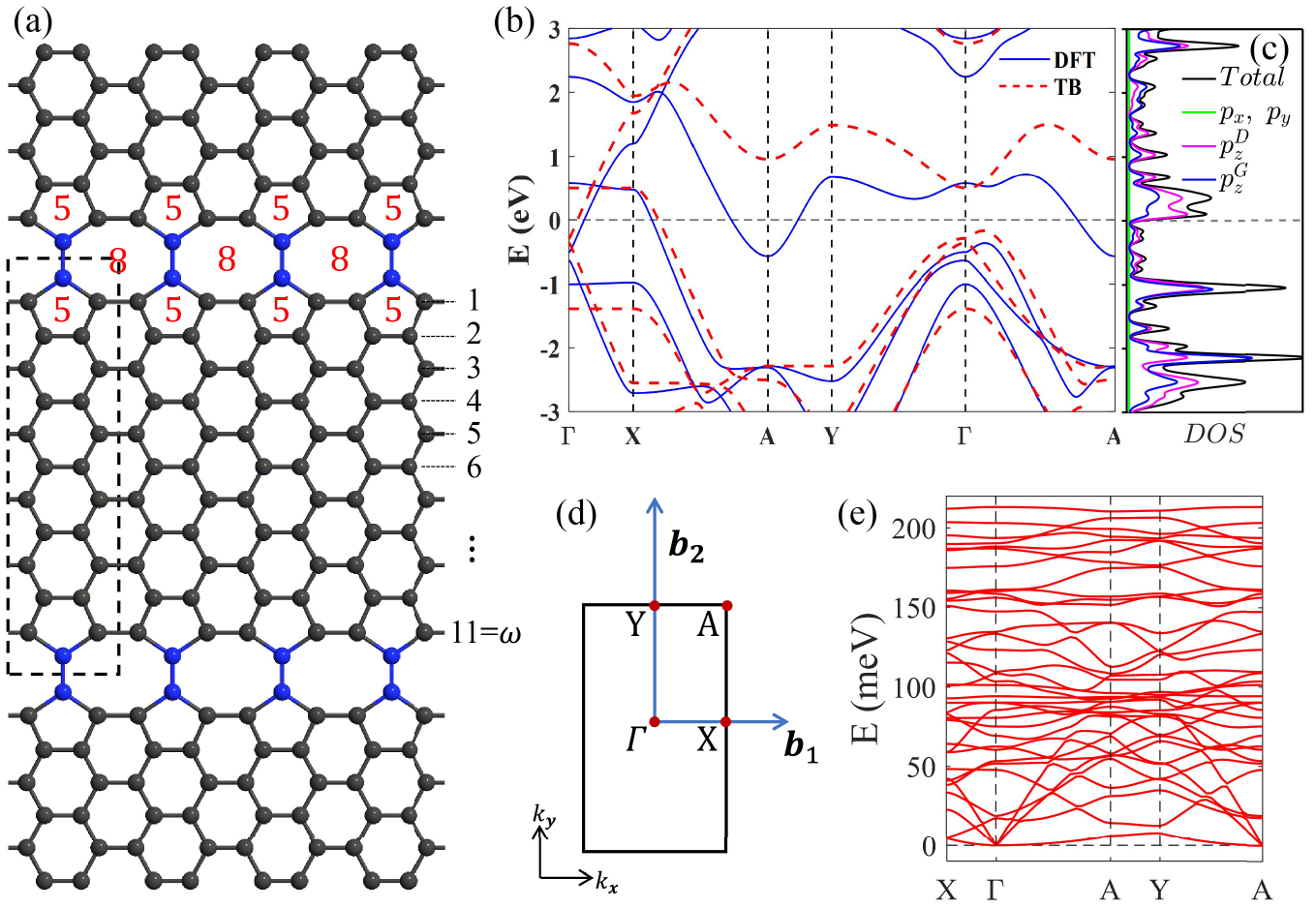}\caption{(a) Structural illustration of $C_{558}$-line defects interspersed in graphene. Two carbon atoms (highlighted in blue) are added along the direction of the armchair graphene nanoribbon, leading to the construction of a line defect structure consisting of two five-membered rings and an eight-membered ring at the interface. The width of the graphene nanoribbons is characterized by the number of rows $\omega$. In this interpolation method, some periodicity is maintained so that the dashed box can be considered as a single cell. (b) Band structure corresponding to $\omega$ = 5, where the blue solid and red dashed lines indicate the results of DFT and TB Calculations; (c) The projected density of states corresponding to (b), where $p_{z}^{D}$ denotes the contribution of carbon atoms near the defect and $p_{z}^{G}$ denotes the orbital projection of the intervening nanoribbon; (d) The first Brillouin zone of the lattice and its high symmetry point. $\boldsymbol{b_1}$ and $\boldsymbol{b_2}$ are two mutually inverse basis vectors; (e) The phonon dispersion relation of the system along the high symmetry point for width $\omega$ = 5.}\label{pho:band}
\end{figure*}

In this work, we study the topological and magnetic properties induced by the $C_{558}$-line defects in graphene, which form five- and eight-membered carbon rings between the subdivided regions. The resulting defect structure induces a topological phase involving a pair of type-\uppercase\expandafter{\romannumeral2} Dirac points near the Fermi energy level. At the same time, we find that the topological properties of the entire system are very sensitive to the coupling strength with line defects. This promotes the occurrence of topological phase transitions. 
We also show that the defect induces a local magnetic moment that spontaneously breaks the time reversal symmetry and leads to a ferromagnetic ground state. By applying an external magnetic substrate, we can regulate the local magnetic moment and achieve spin-polarized transport in nanoribbons of finite width. The energy of the incident electrons can be varied to achieve spin filtering. The topological defect structure reveals the hidden topological property of graphene nanoribbons and their potential for application in spintronic devices.

\section{Calculation methods }

We study the electronic properties of graphene networks with $C_{558}$ line defect based on first-principles methods and tight-binding models. Ab initio software packages used in this article include ATK and QE \cite{36,37}. The generalized gradient approximation (GGA) is used for the exchange-correlation functional. A plane wave basis set with a cutoff energy of 100 Ry and an ultrasoft pseudopotential \cite{38} is used. To account for Coulomb interactions beyond the GGA, we use a newly developed DFT +U+V method that uses self-consistent calculations of onsite and intersite Hubbard interactions (U and V). 

Using the Tight-binding (TB) model is mainly aimed at studying the impact of defects on the topological properties of graphene networks. We use the TB approximation with the $p_z$ orbital to derive all observed properties. Based on the defect structure, we can categorize it into three regions: the region of intermediate graphene nanoribbon (G), the region of line defect (D), and the junction between the defect and the nanoribbon (DG). So, the hopping are defined as $-t_{G}$, $-t_{D}$, and $-t_{ DG }$. These parameters allow us to formulate a TB Hamiltonian
\begin{equation}
{\cal H}_{\mathrm{TB}}=-t\sum_{\langle i,j\rangle}c_{i}^{\dagger}c_{j}+ \sum_{i}\epsilon_{i} c_{i}^{\dagger}c_{i}+\left(\mathrm{c.c.}\right),
\end{equation}
where the first term is the hopping term, $t \in [t_{G}, t_{D}, t_{DG}]$. The second term represents the on-site energy.

The spin transport properties of nanoribbons with $C_{558}$ defects under local exchange magnetic fields are investigated using the nonequilibrium Green's function (NEGF) method. We focus on the ballistic transport state, where there is no inelastic scattering in the channel region. 
Consequently, we can apply the Landauer-B$\rm \ddot{u}$ttiker formula for the spin-dependent conductance as \cite{39}
\begin{equation}
G^s(E)=\dfrac{e^2}{h}Tr\big[\Gamma^s_L(E)g^s(E)\Gamma^s_R(E)g^s(E))^\dagger\big],
\end{equation}
where $G^{s}$ denotes the electron conductance of the spin, $E$ is the electron energy, and $g^s (g^{s\dagger})$ denotes the  retarded (advanced) Green's function matrix. $\Gamma^s_{L(R)}(E)=i\big(\Sigma^s_{L(R)}-\left(\Sigma^s_{L(R)}\right)^\dagger\big) $ is the  broadening matrix between the central region and the left (right) lead of the system. Here, we can calculate numerically the spin-dependent self-energy $\Sigma^s_{L(R)}$ of the left (right) wire by iterating \cite{40}.

\section{Defect structure and electronic properties}
We first introduce the defect structure of two-dimensional graphene monolayers. For the regular graphene, we consider the additional carbon atoms introduced at the armchair-like boundary positions on the real space lattice, leading to the formation of a line defect structure consisting of five- and eight-membered rings at the interface, as shown in Fig. \ref{pho:band}(a) . Although $C_{558}$-line defect breaks the original crystal symmetry of graphene, we still preserve some symmetry by interspersing it with cycles. This is important for the topological properties we will describe in the next section. 

After optimization, for the system with $w = 5$,  the five corners of the pentagonal ring are $109.91^{\circ}$, $101.87^{\circ}$, $113.18^{\circ}$, $113.18^{\circ}$, and $101.87^{\circ}$. Moreover, the carbon-carbon bond length between the two labeled blue carbon atoms at the defect is 1.38 $\AA$, while the bond length between the nearby blue and black carbon atoms is 1.53 $\AA$. The carbon-carbon bond length of the middle black graphene nanoribbon is kept at about 1.43 $\AA$. In two-dimensional carbon-based materials, the hopping displays an exponential relationship with bond length \cite{41}. Therefore, the determination of the hopping parameters in the TB model strongly depends on the bond length. So, we set the hopping in the graphene nanoribbon to $t_{G} = -2.7 eV$, while $t_{D} = 1.2t_{G}$ at the defect and $t_{ DG } = 0.8t_{G}$ at the junction. The band structure obtained from the TB model is consistent with the DFT calculation results, as shown in Fig. \ref{pho:band}(b). Fig. \ref{pho:band}(c) shows the projected density of states, indicating that the $p_z$ orbitals dominate near the Fermi energy level, with the defects showing a more substantial and numerically larger contribution in the $p_z$ orbitals than the graphene nanoribbons. For this defective structure, the primary concern is its stability. We calculate the phonon dispersion relation for this structure at small size, as shown in Fig. \ref{pho:band}(e). The phonon spectrum shows no spurious frequencies, indicating that the system is stable. As the nanoribbon size increases, the effect of the defect on the body decreases and its stability increases. The most important feature is that the formation energy of the system decreases with increasing size and approaches the formation energy of graphene. This result has been discussed and demonstrated for graphene-like interfacial structures \cite{22}.

In electronic structures with $w$ = 5, two Dirac points are located on the high symmetry line between $\Gamma$ and X and between $\Gamma$ and -X, respectively. The Fermi velocities of the energy bands near the Dirac point are far apart, leading to anisotropic linear dispersion except along the $k_{x}$ direction. As the $C_{558}$ line defect embedded system expands, the Dirac dispersion relation exhibits a 3p periodicity, as shown in Fig. \ref{pho:sm} (a-f). The DFT calculation results are detailed in Appendix A. 

\begin{figure}
	\centering
	\includegraphics[width=7.9cm]{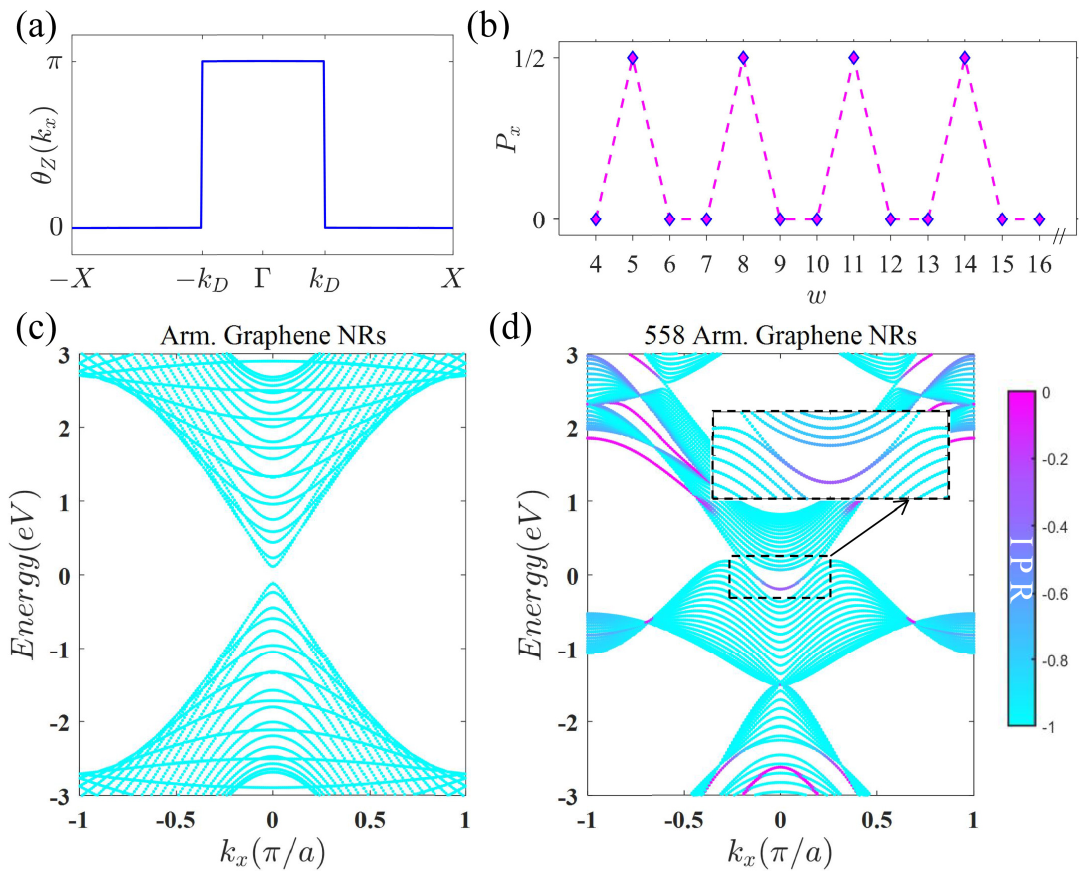}\caption{(a) Zak-Berry phase diagram as a function of $k_{x}$ whre the nanobandwidth $\omega$ is equal to 5. (b) Charge polarization $P_{x}$ as a function of $\omega$ in the $k_x$ direction for defective structures at different scales. (c) Band structure corresponding to the armchair nanoribbon without defects; (d) Band structure interspersed with the nanoribbon with a $C_{558}$ line defect and traversing 15 cycles in the y-direction of real space. A local enlargement of the dashed box is seen near the Fermi energy level.}\label{pho:zak}
\end{figure}

\begin{figure*}
	\centering
	\includegraphics[width=13cm]{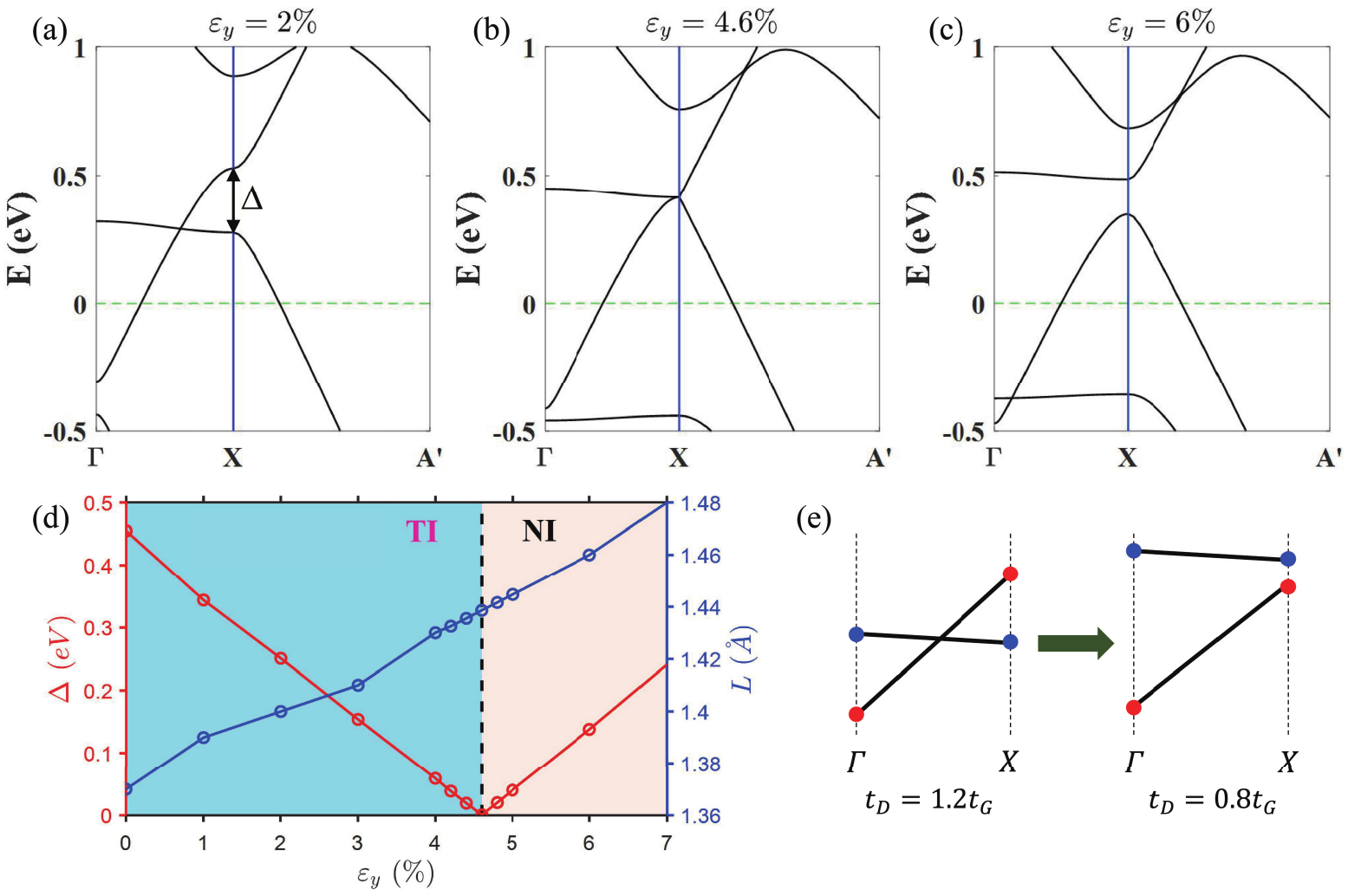}\caption{(a-c) Modulated energy band structure under strains of 2$\%$, 4.6$\%$ and 6$\%$. (d) Variation of $\Delta$ and L with strain, where $\Delta$ represents the energy difference between the unoccupied ($N_{occ}+1$) and occupied ($N_{occ}$) bands at point X, and L represents the bond length between two blue atoms at the line defect. (e) Schematic representation of the TB model for the topological phase transition, where red (blue) denotes even (odd) parity.}\label{pho:tran}
\end{figure*}

\section{Topological properties and phase transitions}

In this section, we demonstrate the topological properties of the system by computing the Zak-Berry phase of the system with the $C_{558}$-line defect. We also discuss in detail the sensitivity of the topological properties of the system to the coupling strength at the defect, in particular with respect to the topological phase transition process after the strain. Since we maintain periodicity of the defects, the system has an effective Brillouin zone in the two-dimensional ($k_x$, $k_y$) direction. In the previous section we learned about its electronic properties, i.e., the appearance of Dirac points on the high $\Gamma-X$ symmetry line, and we expect topological results in this direction. Consequently, we can obtain the Zak-Berry phase only along the ky-integral Berry contact \cite{42}:
\begin{equation}
\theta_{Z}(k_{x})=\int_{-\pi}^{\pi}d k_{y}\mathrm{tr}\mathcal{A}_{y}(k_{y}),
\end{equation}
where
\begin{equation}
\mathcal{A}_{_{y}}(k_{_{y}})=\langle u_{m}(k_{x},k_{_{y}})|i\nabla_{k_{y}}|u_{n}(k_{x},k_{y})\rangle,
\end{equation}
is the Berry connection associated with the occupied states. When the Zak phase is nonzero, the system charge polarizes, producing a topologically protected boundary state. This Zak phase extended to 2D is equivalent to the wave function polarization given below \cite{43}:
\begin{equation}
\mathbf{P}=\frac{1}{2\pi}\int d k_{x}d k_{y}\mathrm{Tr}[\mathbf{A}(k_{x},k_{y})].
\end{equation}
In a two-dimensional crystal system, the charge polarization can be thought of as a vector $P=(P_i, P_j)$ whose component depends on the direction of the wave vector $k_{i(j)}$. For each value of $k_x$, this Zak-Berry phase is quantified by the spatial symmetry $C_{2z}$. The spatial inverse symmetry has a strong constraint on the value of $P$, which is determined independently of the parity of the $\Gamma$ point and the X(Y) point \cite{44}, that is
\begin{equation}
P_{i}=\frac{1}{2}\bigg(\sum_{n}q_{i}^{n}\mathrm{modulo}2\bigg),(-1)^{q_{i}^{p}}=\frac{\eta(X_{i})}{\eta(\Gamma)},
\end{equation}
where $\eta$ is the eigenvalue of the rotation of the energy band along the out-of-plane z-direction, the summation is over all occupied bands, and $i$ represents x or y. 

Figure \ref{pho:zak} illustrates the topological properties of the graphene network structure with $C_{558}$-line defects. Fig. \ref{pho:zak}(a) displays the Zak-Berry phase for the system with $\omega=5$, with $k_x$ ranging from $-\pi$ to $\pi$. The Zak-Berry phase is quantized and changes exactly at the Dirac node $k_D$, indicating its sensitivity to the system's topological properties. This Zak phase extends to 2D, implying that each $k_x$ value represents a one-dimensional insulator oriented along the y-direction, with an effective inversion symmetry generated by $C_{2z}$. For $k_x$ values within the range $[-k_D, k_D]$, each insulator has a Zak phase $\theta_{\text{Z}}=\pi$ and a charge polarization $P_i = \frac{1}{2}$. Conversely, for $k_x$ values outside this range, the bulk exhibits normal behavior with a Zak phase $\theta_{\text{Z}}=0$ and charge polarization $P_i =0$. 
The $C_{2z}$ symmetry operation ensures the degeneracy of defect-localized Dirac nodes and thus protects the system's topological properties. In Fig. \ref{pho:zak}(b), we employ the same $C_{2z}$ symmetry operation to investigate the charge polarization parameter $P_x$ of the multiscale system along the $k_x$ direction. Notably, we find that $P_x= \frac{1}{2}$ when $\omega$ equals $3p + 2$, whereas $P_x$ equals zero for $\omega = 3p$ and $3p+1$. This rule characterizes the band structure properties of various scales along the high-symmetry $\Gamma-X$ line, indicating that periodic $C_{558}$-line defects endow graphene with a three-cycle topological property.

Since our embedding direction is along the armchair boundary, it follows the same pattern as the nanoribbons of graphene with open boundary condition \cite{20}. 
In contrast to ordinary armchair nanoribbons that do not possess the topological characteristic of boundary polarization, the introduction of line defects in our nanoribbons can generate edge states at the armchair boundaries. This is illustrated in the band structure depicted in Fig. \ref{pho:zak}(c,d) for a finite period. The color in the diagram denotes the magnitude of the inverse participation ratio (IPR) \cite{45}, defined as
\begin{equation}
\alpha(E)=\frac{\operatorname{ln}\sum_{j=1}^{M}\left|\psi(j)\right|^{4}}{\operatorname{ln}M},
\end{equation}
where M is the total number of lattice points in the nanoribbon. Its value ranges from $-1$ to 0. The closer to 0, the more localized the wave function is. It well characterizes the edge polarization scale of the wave function at each wave vector position.

\begin{figure}
	\centering
	\includegraphics[width=7cm]{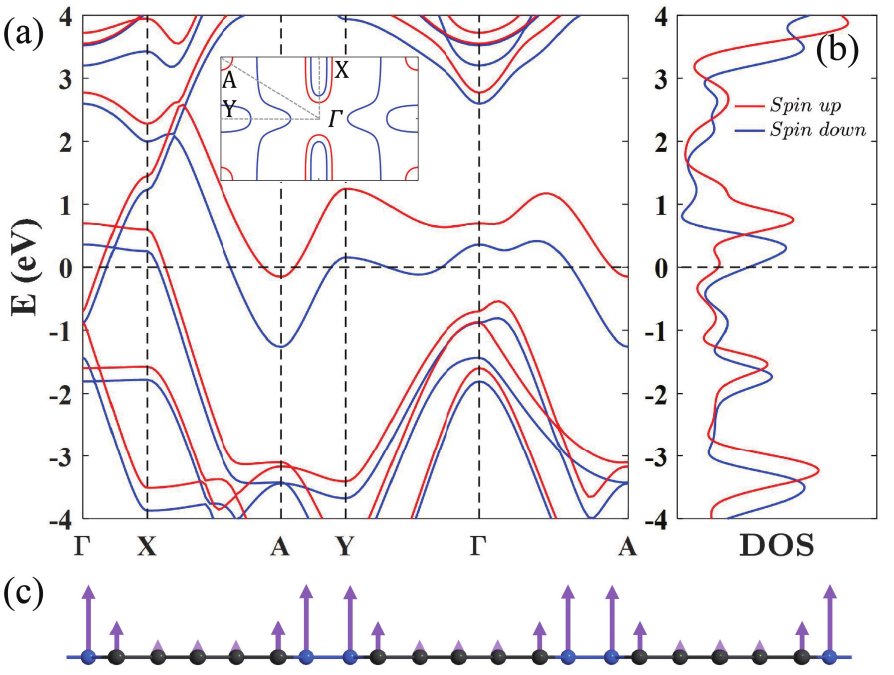}\caption{(a) Band structures obtained by the DFT+U+V method, and the illustration shows the Fermi surface. The red and blue lines indicate the bands with opposite spin directions, and the corresponding spin density of states is shown in (b). (c) Schematic representation of the calculated net magnetic moment at each atomic position.}\label{pho:mag}
\end{figure}

\begin{figure*}
	\centering
	\includegraphics[width=12cm]{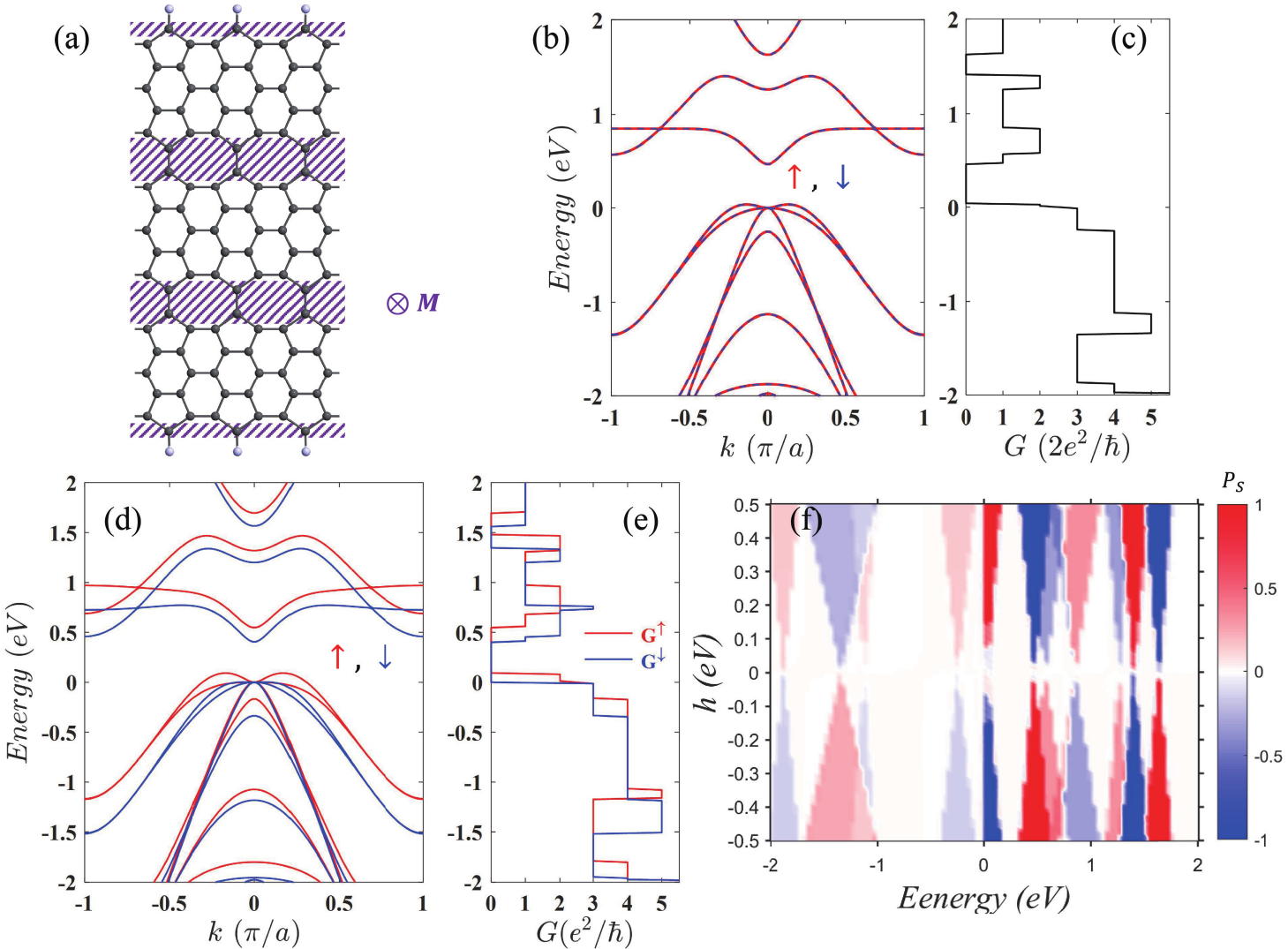}\caption{(a) Structural diagram of the $C_{558}$ defect nanoribbons. The purple shaded area is the local substrate. (b) Band structures of the magnetic moment along the in-plane direction. $\uparrow / \downarrow$ indicate spin-up and spin-down, respectively. (c) Corresponds to the conductivity in b as a function of the energy of the incident electron. (d) The band structures when the substrate magnetic moment is perpendicular to the nanoribbon plane and the local exchange field strength is h = 0.3 eV; (e) Corresponds to the spin-dependent conductance Gup and Gdown in (d). (f) Spin polarization curve as a function of exchange field strength h and incident electron energy E.\label{pho:nano}}
\end{figure*}

Next, we investigate the topological phase transition process caused by stretching to gain a deeper understanding of the coupling between line defects and central nanoribbons. We apply tensile stress only in the y-direction and choose a topological system with $\omega = 8$. As the strain increases, the Dirac point approaches the $X$ point and reaches it at $\varepsilon_y = 4.6 \% $, as shown in Fig. \ref{pho:tran}(b). For more than 4.6 $\%$, the Dirac point disappears with the opening of the band gap, as shown in Fig. \ref{pho:tran}(c). The energy difference ($\Delta$) between the unoccupied ($N_{occ}+1$) and occupied ($N_{occ}$) bands at point X changes with strain, as shown in the red dotted line plot in Fig. \ref{pho:tran}(d). This is a very obvious phase transition process from topological phase to trivial phase. After the strain-optimized structural analysis, we find that the carbon-carbon bond ($L$) at the defect changes the most and increases with strain, as shown in the blue dotted line in Fig. \ref{pho:tran}(d). This has significant implications for the magnitude of $t_D$ in the TB model. At strains close to 4.6$\%$, L is about 1.43 $\AA$, which is comparable to the carbon-carbon bond length of graphene. When the strain exceeds this value, $t_D<t_G$, causing a phase transition. By adjusting $t_D$, the two energy bands associated with the topological phase transition in the TB model shift up and down, resulting in a Zak phase transition from $\pi$ to 0 in the band of the occupied state of $N_{occ}$. The $t_D$ of the topological and trivial phases are taken as $1.2t_G$ and $0.8t_G$, respectively, as shown in Fig. \ref{pho:tran}(e).

In this way, we have gained a comprehensive understanding of the topological periodicity arising from the $C_{558}$-line defect structure. Moreover, our study of the strain-induced topological phase transition process shows how sensitive graphene is to defects and highlights the strong controllability of linear defect structures on system properties. 

\section{local magnetic moments and spin polarization}

The Hubbard-Coulomb interaction, as with other carbon-based materials, can change the energy and slope of the DFT-GGA band. Our calculations, which accounted for the Coulomb interaction beyond GGA, show that the $C_{558}$-line defect graphene network structure is ferromagnetic in its ground state. Our self-consistent calculations, which accounted for Hubbard U+V interactions, yielded a total energy for the magnetic state that is 3.71 meV per atom lower than for the non-magnetic state. For $p_z$ orbitals, U = 6.01 ($\pm$0.05) eV and V = 3.06 ($\pm$0.05) eV are obtained, and these values are relatively small at the defects. These parameters are comparable to those of graphene-like materials \cite{35}. Using a  nanoribbon with a width of $w=5$ as an example, we obtained the spin-splitting band structures as shown in Fig. \ref{pho:mag}(a). The spin polarization occurs at different energy levels, especially near the Fermi level, as shown in Fig. \ref{pho:mag}(b). The calculations give a net magnetic moment of 0.53 $\mu_{\rm B}$ for a single cell, with a local magnetic moment of 0.11 $\mu_{\rm B}$ for the carbon atom at the defect, 0.05 $\mu_{\rm B}$ for the nearest carbon atom, and less than 0.01 $\mu_{\rm B}$ for the remaining atomic positions. Fig. 4(c) shows the distribution of atomic moments, which clearly illustrates the local magnetic moments of graphene caused by line defects. It is important to note that although the nonzero magnetic moment breaks time inversion symmetry, the spatial reflection symmetry is preserved and thus does not affect the topological properties.

While our calculations suggest a ferromagnetic ground state, it may not be spontaneously stable, as seen in other single-layer defect structures \cite{35}, due to in-plane fluctuations. Nevertheless, the addition of an external field or substrate can stabilize and enhance the local magnetic moment at the defect site. Therefore, we investigate the spin-filtering behavior of finite width nanoribbons with a local magnetic moment modulated by the substrate.

To emphasize the impact of line defects on the system, we introduce a magnetic substrate only at the defect site, which stabilizes and strengthens the exchange effect of the magnetic moment at that location. In light of this exchange effect, the Hamiltonian of the system must be written as:
\begin{equation}
\begin{aligned}
H&=\sum_{i,\alpha}\varepsilon_{i\alpha}C^{\dagger}_{i\alpha}C_{i\alpha}
-t\sum_{\left\langle i,j\right\rangle,\alpha}(C^{\dagger}_{i\alpha}C_{j\alpha}+h.c.)\\
&+h\sum_{i,\alpha\in D}C^{\dagger}_{i\alpha}C_{i\alpha}\left[\hat{m}\cdot\hat{\sigma}\right]\text{.} 
\end{aligned}
\end{equation}
The equation presented here introduces a term of the magnetic exchange action term and a spin indicator $\alpha$. The magnetic exchange field with intensity $h$ is caused by the exchange field generated by the nearly ferromagnetic material \cite{46,47}. We chose nanoribbons with a width of three cycles as the transport carrier, as shown in Fig. \ref{pho:nano}(a). We calculate the energy band structure under the influence of an in-plane magnetic exchange field (Fig. \ref{pho:nano}(b)) and find that the spin is not split, indicating that the in-plane exchange field has little effect on the spin. However, it is important to note that there is a nonzero band gap near the Fermi energy level, which is a common feature of semiconductors and cannot be neglected. The corresponding conductance is shown in Fig. \ref{pho:nano}(c). 
The spin-dependent bands of the nanoribbon is shown when the magnetic exchange field is perpendicular to the surface of the substrate. Fig. \ref{pho:nano}(d) shows the spin-split bands for $h$ = 0.3 eV. Furthermore, Fig. \ref{pho:nano}(e) illustrates that the out-of-plane magnetic exchange interaction results in spin polarization, as reflected in the fully polarized conductivity in the energy range of 0.0 eV to 0.1 eV and 0.4 eV to 0.55 eV. This indicates that only electrons with spin down (spin up) can pass through the nanoribbons within these energy intervals. The spin polarization rate is defined as:
\begin{equation}
P_s(E)=\dfrac{G^{\uparrow}-G^{\downarrow}}{G^{\uparrow}+G^{\downarrow}}.
\end{equation}
The spin polarization rate ($P_s$) is limited to the range of [-1,1]. Fig. \ref{pho:nano}(f) shows the contour plot of $P_s$ as a function of magnetic exchange field strength ($h$) and incident electron energy in the range of -2 eV to 2 eV, to study the spin polarization of nanoribbons with line defects. The fully spin-polarized conductance occurs near the Fermi level and at 1.5 eV due to the band gap. However, in the other incident electron ranges, both spin-up and spin-down bands exist, resulting in partial spin polarization since the two bands do not contribute equally to the electron conductivity. This suggests that the local magnetic moment at the defect location can modulate the perfect spin polarization. By applying the localized magnetic exchange field, graphene nanoribbons interspersed with $C_{558}$-line defects can act as spin filters, tuning the spin direction of the transferred electrons by adjusting the incident electron energy.

\section{CONCLUSIONS}

In summary, we have successfully developed a new topological material by introducing $C_{558}$-line defects into a monolayer of graphene. For this graphene network, its topological phase is protected by $C_{2z}$ symmetry and exhibits a triperiodic pattern. Moreover, the topological properties of this system are highly sensitive to defects, allowing for topological phase transitions by modulating the hopping of electrons at the defect sites through strain. At the same time, we discuss the magnetism and spin effects induced by $C_{558}$ line defects, which expands the selection and scope of spin-based materials. We theoretically prove the ferromagnetic ground state and spin polarization of the embedded system through computational analysis. The local magnetic moments primarily concentrate at defect sites. Finally, by utilizing a magnetic substrate to modulate the magnetic moments at defect sites, we have achieved tunable spin-polarized transport in the limited nanoribbon system. We propose achieving flawless spin filtering by utilizing nanoribbon-based devices.

\section*{ACKNOWLEDGMENTS}

We acknowledge the financial support by the National Natural Science Foundation of China (No. 11874113) and the Natural Science Foundation of Fujian Province of China (No. 2020J02018).

\section*{APPENDIX : THE 3P RULE OF BAND STRUCTURES}
\begin{figure}
	\centering
	\includegraphics[width = 8cm]{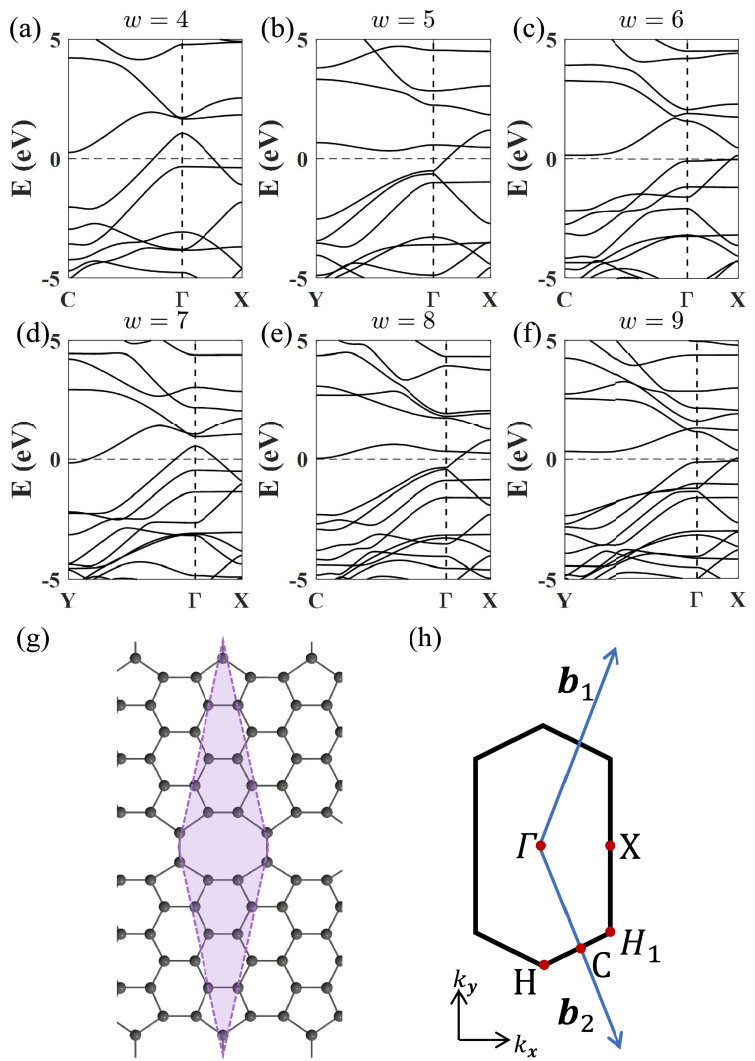}\caption{(a-f) Band structures corresponding to $\omega = 4 \sim 9$ . (g) If $w$ is an even number, the form of a single cell; And the first Brillouin zone and its high symmetry point, $b_1$ and $b_2$ are two reciprocal basis vectors appearing in (h).}\label{pho:sm}
	
\end{figure}

In this section, we introduce the periodic band structures related to the nanoribbon width, as shown in  Fig. \ref{pho:sm}. Although we show the bands for only two periods, it is sufficient to observe the dispersion 
pattern of the bands within the $\Gamma-X$ interval. When $w = 3p+2$, the pair of Dirac points appears in the $k_x$ direction, while this property is not present for other widths of nanoribbons. It is worth noting that when the width $w$ characterizing the nanoribbon is even, the symmetry of the system is 
different and the unit cell is rhombic, as shown in Fig. \ref{pho:sm}(h). However, this difference does not affect the properties of the system in the $k_x$ direction. This periodicity is identical to that of the 
armchair nanoribbons. The reason for this is due to the interleaved $C_{558}$-line defects, which break the original periodicity and add a full wavelength, so that this armchair interface condition also exhibits three-periodicity in the defect structure, as is the case for other similar materials \cite{22}.

\bibliography{References-fop.bib}


\end{document}